\documentclass[authoryear]{FLO_v1}%

\usepackage{setspace}
\usepackage{graphicx}
\usepackage{import}
\usepackage{upgreek}
\usepackage{amsmath,amssymb,amsfonts}
\usepackage{mathrsfs}
\usepackage{amsthm}
\usepackage[figuresright]{rotating}
\usepackage{appendix}
\usepackage[authoryear]{natbib}
\usepackage{ifpdf}
\usepackage[T1]{fontenc}
\usepackage{newtxtext}
\usepackage{newtxmath}
\usepackage{textcomp}
\usepackage{xcolor}
\usepackage{transparent}

\usepackage[colorlinks,allcolors=blue]{hyperref}
\definecolor{jourcolor}{cmyk}{1,0.57,0.01,0.38}
\hypersetup{
    colorlinks,%
    citecolor=jourcolor,%
    filecolor=jourcolor,%
    linkcolor=jourcolor,%
    urlcolor=jourcolor
}

\usepackage{commath,physics}
\usepackage[nameinlink,noabbrev]{cleveref}%

\reversemarginpar
\usepackage[color=red!80!black, textcolor=white, textsize=small]{todonotes}

\usepackage[locale=UK,
number-mode=math,
unit-mode=text,
per-mode=symbol,
number-unit-product=\ ,
retain-zero-exponent=false]{siunitx}[=v2]
\DeclareSIUnit{\pixel}{px}
\DeclareSIUnit{\fps}{fps}

\newcommand{\kindex}[2]{\ensuremath{{#1}_{\scalebox{0.65}{#2}}}}
\newcommand{\U}{\textrm{U}}
\newcommand{\Uinf}{\kindex{\U}{$\infty$}}

\newcommand{\Rey}{\mbox{Re}}
\newcommand{\subf}[1]{(#1)}

\graphicspath{{../../figurematter/latex/figs/}{../../figurematter/plots/}}

\articletype{PREPRINT}

\usepackage{xcolor, tikz}
\usetikzlibrary{shapes, patterns}
\usepackage{xspace}

\definecolor{epflblue}{RGB}{31,55,91}
\definecolor{epflred}{RGB}{255,0,0}
\definecolor{groseille}{RGB}{181,31,31}
\definecolor{leman}{RGB}{0,167,159}
\definecolor{canard}{RGB}{0,116,128}
\definecolor{ardoise}{RGB}{65,61,58}
\definecolor{perle}{RGB}{202,199,199}
\definecolor{taupe}{RGB}{69, 58, 76}
\definecolor{montrose}{RGB}{242, 151, 105}
\definecolor{vertdeau}{RGB}{193, 220, 175}
\definecolor{rose}{RGB}{236, 110, 155}
\definecolor{acier}{RGB}{79, 142, 203}
\definecolor{souffre}{RGB}{251, 237, 102}
\definecolor{carotte}{RGB}{235, 102, 8}
\definecolor{zinzolin}{RGB}{92, 36, 130}
\definecolor{chartreuse}{RGB}{127, 255, 0}
\definecolor{marron}{RGB}{91, 52, 40}
\colorlet{stage1}{rose!50}
\colorlet{stage2}{leman}
\colorlet{stage3}{canard}
\colorlet{stage4}{zinzolin}
\colorlet{stage5}{acier}
\colorlet{stage6}{acier!50}

\newcommand{\stage}[1]{\raisebox{0pt}{\xspace\tikz{\draw[fill=#1!50, draw=#1, line width=1pt] (0,0) rectangle +(2ex,1ex);}\xspace}}

\newcommand{\Cppos}{\raisebox{0pt}{\xspace\tikz{\draw[pattern=north west lines, pattern color=vertdeau, draw=ardoise, line width=1pt] (0,0) rectangle +(2.48ex,1.5ex);}\xspace}}
\newcommand{\Cpneg}{\raisebox{0pt}{\xspace\tikz{\draw[pattern=north east lines, pattern color=montrose, draw=ardoise, line width=1pt] (0,0) rectangle +(2.48ex,1.5ex);}\xspace}}

\begin{document}

\title{Timescales of dynamic stall development on a vertical-axis wind turbine blade}
\author{S\'ebastien Le Fouest, Daniel Fernex, and Karen Mulleners$^{\ast}$}%
\address{Institute of Mechanical Engineering, \'Ecole Polytechnique F\'ed\'erale de Lausanne (EPFL), CH-1015 Lausanne, Switzerland}
\corres{*}{Corresponding author. E-mail:\emaillink{karen.mulleners@epfl.ch}}


\abstract{
Vertical-axis wind turbines are great candidates to diversify wind energy technology, but their aerodynamic complexity limits industrial deployment. 
To improve the efficiency and lifespan of vertical axis wind turbines, we desire data-driven models and control strategies that take into account the timing and duration of subsequent events in the unsteady flow development.
Here, we aim to characterise the chain of events that leads to dynamic stall on a vertical-axis wind turbine blade and to quantify the influence of the turbine operation conditions on the duration of the individual flow development stages. 
We present time-resolved flow and unsteady load measurements of a wind turbine model undergoing dynamic stall for a wide range of tip-speed ratios. 
Proper orthogonal decomposition is used to identify dominant flow structures and to distinguish six characteristic stall stages: the attached flow, shear-layer growth, vortex formation, upwind stall, downwind stall, and flow reattachment stage. 
The timing and duration of the individual stages are best characterised by the non-dimensional convective time. 
Dynamic stall stages are also identified based on aerodynamic force measurements. 
Most of the aerodynamic work is done during the shear-layer growth and the vortex formation stage which underlines the importance of managing dynamic stall on vertical-axis wind turbines.
}

\maketitle

\begin{boxtext}

\textbf{\mathversion{bold}Impact Statement}

Wind energy can play a crucial role in achieving a zero-emission power grid by 2050. One way to increase the installed wind energy capacity is by diversifying wind turbine technologies. Vertical-axis wind turbines are ideal for urban and floating offshore applications, and can complement traditional turbines to achieve higher power density wind farms.
Large-scale deployment of vertical-axis wind turbines has been challenging due to their aerodynamic complexity and the occurrence of dynamic stall on the turbine blades.
Active flow control can alter the occurrence of dynamic stall on the turbine blade and mitigate load transients.
This work aims to characterise the start and duration of landmark dynamic stall stages to facilitate the design of robust, physics-based control laws to improve the performance of vertical-axis wind turbines.
The ability to detect transitions between different stages in the flow development using only force measurements is validated.

\end{boxtext}


\section{Introduction}
\label{sec:intro}


Wind power has the potential to play a significant role in reducing the carbon intensity of the global power grid and in reaching the Paris Agreement goal to limit global warming below \SI{2}{\degreeCelsius} \citep{UNFCCC.ConferenceofthePartiesCOP2015}.
A sound strategy to increase the installed wind energy production capacity is to diversify wind energy extraction technologies \citep{Miller2021,Rolin2018,Jamieson2011}.
Vertical-axis wind turbines feature many advantages that make them attractive for wind energy exploitation in areas where their horizontal counterparts may face shortcomings, such as urbanised regions or floating offshore applications \citep{DeTavernier2019,Rezaeiha2018a,Nguyen2017}.
These advantages include low noise emissions, high power densities, and insensitivity to wind direction \citep{Buchner2018,SimaoFerreira2009}.
Vertical axis wind-turbines have typically fewer mechanical components than horizontal axis turbines and heavy components such as the drive train and the generator can be placed closer to the ground which facilitates commissioning and maintenance.
The performance of wind farms can also be significantly improved by taking advantage of the complementarity of vertical-axis wind turbines and traditional wind turbines \citep{Dabiri2011,Strom2022,Wei2021}.

\begin{figure}
\centering
\includegraphics{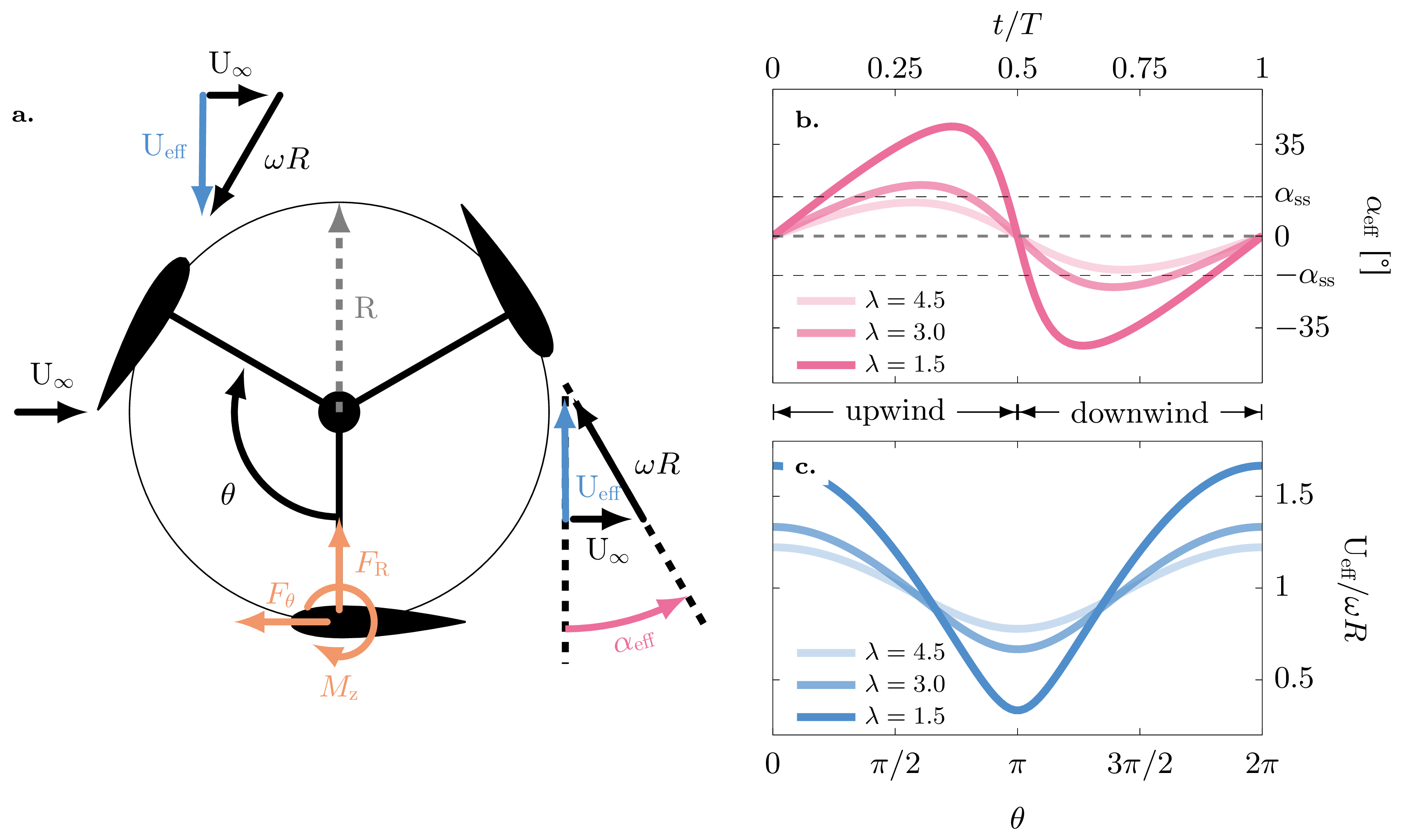}
\caption{Vertical-axis wind turbine blade kinematics.
\subf{a} The free stream velocity $\protect\Uinf$ goes from left to right.
Indication of the positive direction of the radial force \kindex{F}{R}, azimuthal force \kindex{F}{$\theta$}, and pitching moment at quarter-chord \kindex{M}{z}.
The blade's velocity is equivalent to the rotational frequency $\omega$ times the turbine's radius $R$.
Schematic representation of the definition of the blade's effective angle of attack $\kindex{\alpha}{eff}$ and velocity $\kindex{U}{eff}$ and their temporal evolution (respectively in \subf{b} and \subf{c}) as a function of the blade's azimuthal position.
}
\label{fig:vawtaero}
\end{figure}

The development of vertical-axis wind turbines was limited by the complexity of the blade aerodynamics and the resultant lower performance compared to their horizontal counterparts \citep{Rezaeiha2017,Ferreira2009,Hwang2009}.
Even with steady inflow conditions, the blade inherently undergoes unsteady kinematics.
The blade kinematics are often described as the summation of unsteady pitching and surging terms.
This description results from a geometric analysis of the problem.
The effective flow conditions are expressed using a vectorial sum of the free-stream velocity $\Uinf$ and the blade velocity $\kindex{U}{b} = \omega R$, where $\omega$ is the turbine rotation frequency, and $R$ is the turbine radius (\cref{fig:vawtaero}).
When we consider an azimuthal position $\theta = \ang{0}$ when the blade faces the wind, we can write the variation of the effective velocity as a function of its azimuthal position $\theta$ as
\begin{equation}
	\kindex{\U}{eff}(\theta) = \sqrt{1+2 \lambda \cos \theta+\lambda^{2}}
	\label{eq:Ueff}
\end{equation}
and the variation of the effective angle of attack as
\begin{equation}
	\kindex{\alpha}{eff}(\theta) = \tan^{-1}\left(\frac{\sin \theta}{\lambda+\cos \theta}\right)
	\label{eq:Aeff}
\end{equation}
where $\lambda = \kindex{U}{b}/\Uinf$ is the tip-speed ratio.
The tip-speed ratio is shown to govern the amplitude of the periodic variation in effective flow conditions in \cref{eq:Aeff,eq:Ueff}.
For low and medium tip-speed ratios, typically $\lambda < 4.5$, the effective angle of attack exceeds the blade's critical stall angle twice during the rotation, which can lead to flow separation on the blade surface.

The severity of flow separation events increases for decreasing tip-speed ratios.
The tip-speed ratio is considered low when the turbine blade exceeds its static stall angle for longer than \num{4.5} convective times, corresponding to the minimum amount of time required for a coherent leading-edge vortex to form \citep{Dabiri2009, LeFouest2021, LeFouest2022}.
Vertical-axis wind turbines operating at low tip-speed ratios ($\lambda \leq 2.5$ when $c/D = 0.2$) undergo deep dynamic stall, which is characterised by the formation, growth, and shedding of large-scale coherent vortices \citep{Carr1977}.
The shedding of large-scale vortices is generally followed by a dramatic loss in aerodynamic efficiency and highly unsteady loads that can potentially lead to turbine failure \citep{Mccroskey1982,Mulleners2013,Corke2015a}.

The temporal occurrence and landmark stages characterising dynamic stall on two-dimensional pitching airfoils in a steady free-stream flow are well documented \citep{Carr1977,Degani1998,Mulleners2013} and their load responses can be reasonably well modelled \citep{Leishman1989,Goman1994,Ayancik.2022}.
For a rotating wind turbine blade, the occurrence of dynamic stall is affected by the circularity of the blade's path in several ways.
The dynamic stall vortex is convected downstream along the blade's path after it is shed, extending the blade-vortex interaction compared to a blade pitching in a steady flow.
The flow topology is dominated by the extended presence of a dissipating dynamic stall vortex at the overlap between the upwind and downwind halves of the blade rotation when the effective velocity is at its lowest (\cref{fig:vawtaero}) \citep{Ferreira2009,Dave2021a}.
The flow and wake curvature also influence the effective flow acting on a vertical-axis wind turbine blade, leading to virtual camber and incidence effects \citep{Migliore1980, Benedict2016}.
The blade crosses the same stream tube twice, leading to repeated blade-wake interactions even for a single-blade configuration \citep{Fujisawa2001, Paraschivoiu1988}.
Typical descriptions of effective flow conditions on a turbine blade, based on \cref{eq:Aeff,eq:Ueff} do not faithfully capture the complexity of the events unfolding during the blade's rotation, even when coupled with an additional induced velocity term \citep{Paraschivoiu1983, Ayati2019}.

Low-order models, such as the actuator cylinder model, improve the prediction of effective flow conditions and unsteady loads by accounting for the velocity induced to the flow by the presence of a rotating turbine \citep{Madsen1982}.
The absence of a reliable dynamic stall model for vertical-axis wind turbines limits the validity of low-order models for low to intermediate tip-speed ratios \citep{Dave2021}.
The most widely used dynamic stall models were developed for pitching blades operating in a free stream \citep{Leishman1989, Goman1994}.
These models rely on a sound understanding of landmark timescales that govern the development of flow separation, namely the vortex formation time and vortex shedding frequency \citep{Ayancik.2022}.
For a vertical-axis wind turbine, the timescales of landmark stall events, such as shear layer roll-up or stall onset, have not yet been characterised.
Dynamic stall timescales are crucial parameters in the overall design and modelling of a vertical-axis wind turbine and for more advanced applications such as devising active flow control strategies.

This investigation aims at improving our qualitative and quantitative understanding of the chain of events leading to dynamic stall on a vertical-axis wind turbine blade and the associated timescales.
We present time-resolved velocity field and load measurements on a scaled-down H-type Darrieus wind turbine operating in a water channel.
A large range of tip-speed ratios is covered to yield a comprehensive characterisation of the occurrence of deep stall on a vertical-axis wind turbine.

\section{Experimental materials and methods}
\label{sec:exp}
The experiments were conducted in a recirculating water channel with a test section of \SI{0.6x0.6x3}{\meter} and a maximum flow velocity of \SI{1}{\meter\per\second}.
A cross-sectional view of the full experimental apparatus is shown in \cref{fig:expsetup}.
The main features of the experimental set-up, force and velocity field measurements, and data analysis using proper orthogonal decomposition are summarised here. 
More details can be found in the supplementary material. 

\subsection{Experimental set-up}

\begin{figure}[tb!]
\centering
\includegraphics{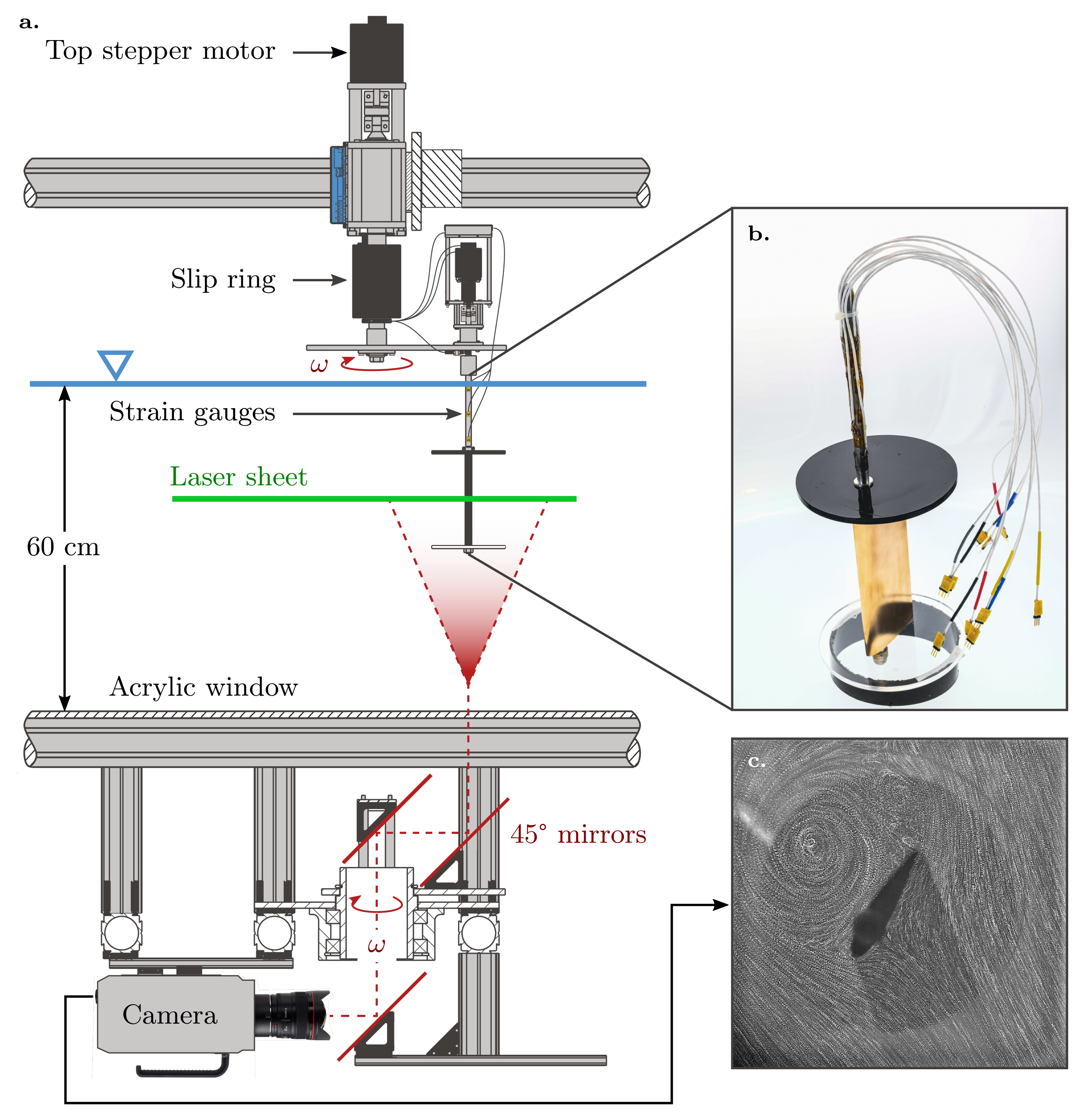}
\caption{\subf{a} Cross-sectional view of the experimental set-up including the wind turbine model, the light sheet, the rotating mirror system, and the high-speed camera for particle image velocimetry. \subf{b} A close-up view of the blade sub-assembly, with installed strain gauges. \subf{c} The camera's field of view indicated by a long exposure image of seeding particles in the flow. }
\label{fig:expsetup}
\end{figure}

A scaled-down model of a single-bladed H-type Darrieus wind turbine was mounted in the centre of the test section (\cref{fig:expsetup}a).
The turbine has a variable diameter $D$ that was kept constant at \SI{30}{\centi\meter}.
Here, we used the single-blade configuration to focus on the flow development around the blade in the absence of interference from the wakes of other blades.
The blade has a NACA0018 profile with a span of $s=\SI{15}{\centi\meter}$ and a chord of $c=\SI{6}{\centi\meter}$, yielding a chord-to-diameter ratio of $c/D=\num{0.2}$.

The turbine model is driven by a NEMA 34 stepper motor with a \ang{0.05} resolution for the angular position.
The rotational frequency was kept constant at \SI{0.89}{\hertz} to maintain a constant chord-based Reynolds number of $\kindex{\Rey}{c} = (\rho \omega R c)/\mu = \num{50000}$, where $\rho$ is the density and $\mu$ the dynamic viscosity of water.
To investigate the role of the tip-speed ratio in the occurrence of dynamic stall, we systematically vary the water channel's incoming flow velocity from \SIrange{0.14}{0.70}{\meter\per\second} to obtain tip-speed ratios ranging from \numrange{1.2}{6}.

\subsection{Force measurements}

The aerodynamic forces acting on the turbine blade are recorded at \SI{1000}{\hertz} for \num{100} full turbine rotations using strain gauges that are glued on to the shaft (\cref{fig:expsetup}b).
The forces presented in this paper are the two shear forces applied at the blade's mid-span in the radial \kindex{F}{R} and azimuthal \kindex{F}{$\theta$} direction, and the pitching moment about the blade's quarter-chord \kindex{M}{1/4} (\cref{fig:vawtaero}).
The total force applied to the blade is computed by combining the two shear forces: $\kindex{F}{tot} = \sqrt{\kindex{F}{R}^2 + \kindex{F}{$\theta$}^2}$.
All force coefficients are non-dimensionalised by the blade chord $c$, the blade span $s$, and the blade velocity $\kindex{\U}{b} = \omega R$ such that:
\begin{equation}
\kindex{C}{tot/R/$\theta$} = \frac{\kindex{F}{tot/R/$\theta$}}{0.5\rho \kindex{\U}{b}^2 sc}\quad.
\end{equation}
The subscripts tot, R, or $\theta$ refer to the total force, the radial, or the tangential force component.

From our load measurements, we retrieve an idealised turbine power coefficient.
The instantaneous power $P$ generated by a vertical-axis wind turbine is linearly proportional to the tangential aerodynamic force experienced by the blade:
\begin{equation}
P(\theta) = \kindex{F}{$\theta$}\, \omega R ,
\end{equation}
where $\omega$ is the rotational frequency and $R$ is the turbine's radius.
We can calculate the instantaneous power coefficient by normalising the estimated instantaneously generated power by the power available in the flow:
\begin{equation}
\kindex{C}{P}(\theta) = \frac{P}{0.5\rho \Uinf^3 \kindex{A}{swept}} \quad,
\end{equation}
where \kindex{A}{swept} is the turbine's swept area given by the product of the turbine diameter and the blade's span.
This definition of the power coefficient is idealised because it does not account for any mechanical or electronic losses that would occur on a wind turbine between the turbine blade torque generation and the generator's output.

\subsection{Particle image velocimetry}

In addition to the blade load measurements, we conducted time-resolved particle image velocimetry (PIV) in the mid-span cross sectional plane of the blade with an acquisition frequency of \SI{1000}{\hertz}.
With the help of a rotating mirror system that spins with the same frequency as the wind turbine, we were able to keep a field of view of \SI{2.5x2.5}{c} that is centred around the blade (\cref{fig:expsetup}c).

The particle images were processed following standard procedures using a multi-grid algorithm \citep{Raffel2007}.
The final window size was \SI{48x48}{\pixel} with an overlap of \SI{75}{\percent}.
This yields a grid spacing or physical resolution of $\SI{1.7}{\milli\meter} = 0.029c$.
A window overlap above \SI{50}{\percent} was selected to minimise spatial averaging of the velocity gradients by the interrogation window following \cite{Richard.2006, kindlerAperiodicityFieldFullscale2011}.
The out-of-plane vorticity component was calculated from the in-plane velocity components using a central difference scheme.  

The experimental facility and apparatus were previously described by the authors in \cite{LeFouest2022}.
More details on the experimental procedure and the turbine model are provided in the supplementary material.

\subsection{\label{subsec:POD} Proper orthogonal decomposition}

The flow development during a full blade rotation is characterised by significant changes in the flow topology and the formation of vortices when dynamic stall occurs.
To identify energetically relevant flow features and their time evolution, we will apply a proper orthogonal decomposition (POD) of the vorticity field.
The POD method was introduced to the field of fluid mechanics to identify coherent structures in turbulent flows \citep{Lumley2007Book}.
The spatial modes \kindex{\psi}{n} reveal the dominant and recurring patterns in the data.
The corresponding temporal coefficients \kindex{a}{n} indicate the dynamic behaviour of the coherent spatial patterns identified in the modes and can be used to extract the characteristic physical timescales of the problem.
As the flow under consideration is vortex dominated, we have opted to decompose the vorticity field instead of the velocity field which was used by \cite{Dave.2023}.
The vorticty field highlights both the rotational and the strain-dominsted regions which are both important to understand dynamic stall development \citep{Menon.2021}.
The POD decomposes the vorticity snapshots at time \kindex{t}{i} into a sum of orthonormal spatial modes $\kindex{\mathbf{\psi}}{n}(x, y)$ and their time coefficients $\kindex{a}{n}(\kindex{t}{i})$ such that
\begin{equation}
	\mathbf{\omega}(x, y, \kindex{t}{i})= \sum\limits_{n=1}^{N}\kindex{a}{n}(\kindex{t}{i})\kindex{\psi}{n}(x, y),
	\label{eq:POD}
\end{equation}
where $N$ is the total number of snapshots.
The POD modes \kindex{\psi}{n} form an optimal basis, from which the original system can be reconstructed with the least number of modes compared to any other basis \citep{Sirovich1987QAM, Taira2017}.

In this study, we have used the phase-averaged snapshots of the flow field for the PIV analysis to focus on the dynamics within a rotational cycle and to decrease the computational cost of the POD.
The dominant POD modes and large-scale vortical structures are similar with both the time-resolved and phase-averaged data.
More information about the phase-averaging procedure are included in the supplementary material. 
The region of interest for the POD analysis covers the entire overlapping measurement field of view which covers the airfoil and the main flow features on each side. 

A key aspect of our study is to compare the dominant flow features shared across varying tip-speed ratios and their relative importance for each case.
To this purpose, we have stacked five series of $200$ phase-averaged vorticity fields $\omega^{\lambda}(x,y,\kindex{t}{i})$, with $i=1,\ldots,200,$ corresponding to five tip-speed ratios $\lambda \in
[1.2, 1.5, 2, 2.5, 3]$ along the time axis into a single matrix $\Omega =
\left[\omega^{\lambda={1.2}}, \omega^{\lambda={1.5}}, \ldots,
\omega^{\lambda={3}}\right]^T$ containing $1000$ snapshots.
The POD of the combined matrix $\Omega$ yields the spatial modes $\kindex{\psi}{n}(x,y)$, which are common for all cases.
The mode coefficient matrix comprises the mode coefficients corresponding to the individual cases $\boldmath{A}=\left[a^{\lambda=1.2}, a^{\lambda=1.5}, \ldots, a^{\lambda=3}\right]$.
The POD eigenvalues $\kindex{\zeta}{n}$ indicate the contribution of each mode $n$ to the total energy of the stacked vorticity fields.
They are strictly decreasing as POD ranks the modes according to their energy content.
With this approach, the vorticity fields for each tip-speed ratio can be reconstructed based on the same set of spatial modes.
The temporal contribution of these common structures to individual tip-speed ratios $\lambda$ is given by the corresponding temporal coefficients $\kindex{a}{n}^{\lambda}$.
A vorticity field for a specific $\lambda=\kindex{\lambda}{m}$ can be reconstructed as
\begin{equation}
	\mathbf{\omega}^{\lambda=\kindex{\lambda}{m}}(x, y, \kindex{t}{i})=\sum\limits_{n=1}^{N}\kindex{a}{n}^{\lambda=\kindex{\lambda}{m}}(\kindex{t}{i})\kindex{\psi}{n}(x,y)\quad.
\end{equation}
The energy contribution $\kindex{\zeta}{n}^{\lambda=\kindex{\lambda}{m}}$ of individual spatial modes $\kindex{\psi}{n}(x,y)$  in the tip-speed ratio case $\lambda=\kindex{\lambda}{m}$ is computed from the corresponding time coefficients as
\begin{equation}\label{eq:zeta}
\kindex{\zeta}{n}^{\lambda=\kindex{\lambda}{m}}=\left<\kindex{a}{n}^{\lambda=\kindex{\lambda}{m}},\kindex{a}{n}^{\lambda=\kindex{\lambda}{m}}\right>=\frac{1}{\kindex{N}{t}}\sum\limits_{i=1}^{\kindex{N}{t}} \left(\kindex{a}{n}^{\lambda=\kindex{\lambda}{m}}(\kindex{t}{i})\right)^2\quad ,
\end{equation}
where $\kindex{N}{t}=200$ is the number of phase-averaged snapshots of the individual tip-speed cases.
The procedure used here is similar to the parametric modal decomposition presented by \cite{Coleman:2018ky}.


\section{Results}
\label{sec:res}


\subsection{\label{subsec:OV} Dynamic stall overview}

\begin{figure}[tb]
\centering
\includegraphics[, trim=0 10mm 0 10mm,clip]{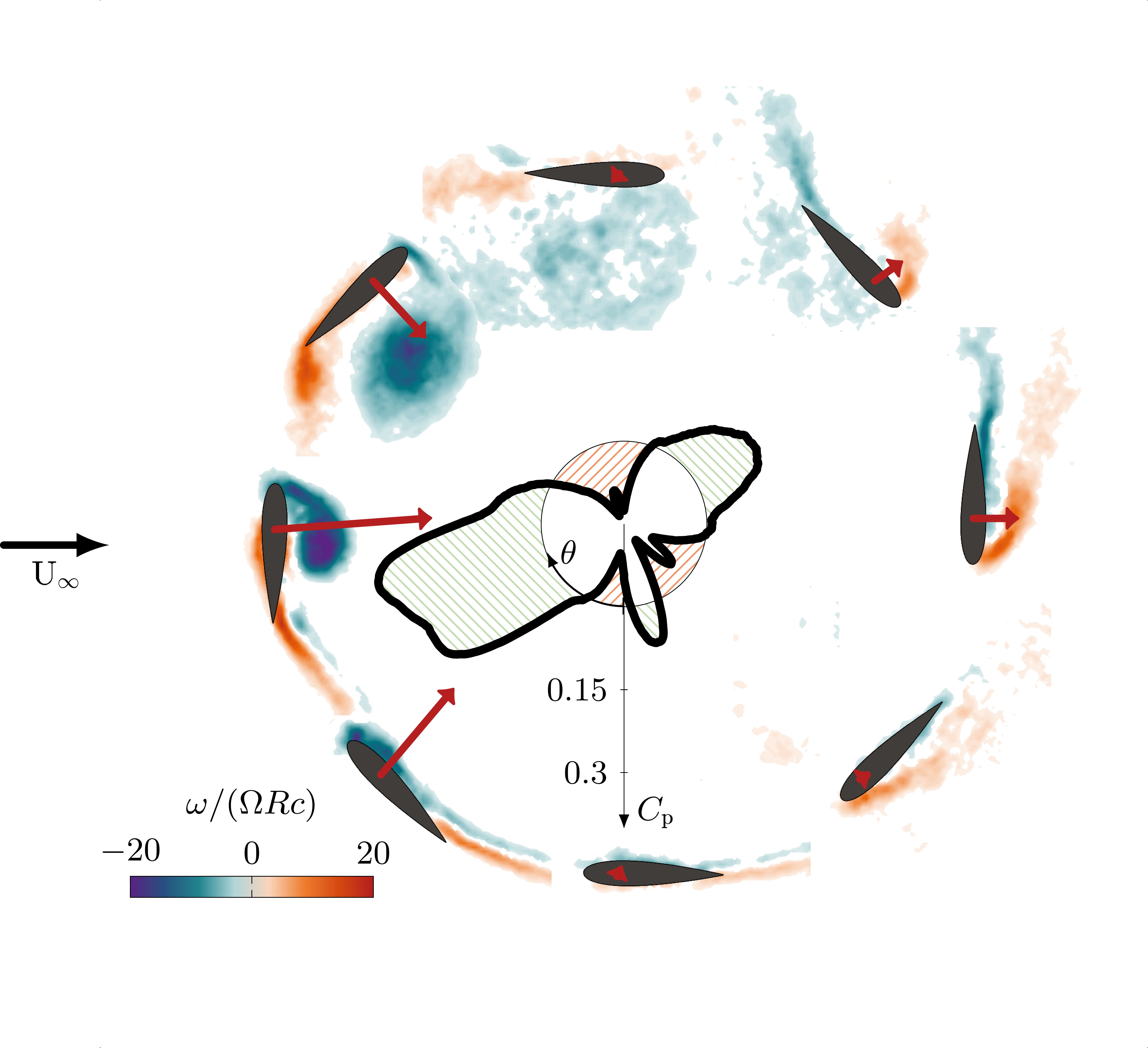}
\caption{Polar plot of the temporal evolution of the phase-averaged power coefficient \kindex{C}{P} for a vertical axis wind turbine operating at tip-speed ratio $\lambda = 1.5$.
Phase-averaged normalised vorticity fields are shown throughout the blade's rotation.
The total aerodynamic force acting on the blade at the various azimuthal locations is depicted by arrows starting from the blade's quarter-chord.
The length of the arrows indicates the relative magnitude of the force.
The black circle indicates $\kindex{C}{p}=0$.
The hatched regions \protect\Cppos and \protect\Cpneg respectively represent regions where the blade experiences net thrust or net drag.
}
\label{fig:res_DS}
\end{figure}

We first present an overview of the development of flow structures and aerodynamic forces experienced by a vertical-axis wind turbine blade undergoing deep dynamic stall.
The temporal evolution of the phase-averaged power coefficient for a wind turbine operating at tip-speed ratio $\lambda = 1.5$ is shown as a polar plot in \cref{fig:res_DS}.
The interplay between the power coefficient and the flow structures forming around the wind turbine blade is highlighted with phase-average vorticity fields presented throughout the blade's rotation.
The total aerodynamic force orientation and relative magnitude are represented by an arrow starting from the blade's quarter-chord.

When the blade is facing the wind ($\theta =$ \ang{0}), the flow is attached and vorticity is only present in the blade's wake.
The total force is small but increases rapidly when the blade moves upwind.
At $\theta = \ang{33}$, the blade's effective angle of attack exceeds its critical static stall angle, which was measured to be $\kindex{\alpha}{ss}=\ang{13}$, and flow reversal emerges on the suction side of the blade.
The shear layer rolls up to form a coherent dynamic stall vortex at the leading edge of the blade.
The presence of this leading-edge dynamic stall vortex causes the power coefficient to rise from \num{-0.03} at $\theta = \ang{0}$ to its maximum value of \num{0.32} at $\theta = \ang{77}$.
Thereafter, the dynamic stall vortex continues to grow in size and strength and it moves from the leading edge towards the mid-chord position.
This migration redirects the aerodynamic force in the radial direction and the power coefficient rapidly drops when the vortex detaches from the blade around $\theta = \ang{114}$.
After lift-off, the stall vortex is convected downstream along the blade's path, dissipates, and breaks apart into smaller scale structures.
The massive flow separation leads to a drop in the total forces acting on the blade and a collapse of the power production.
The power coefficient remains around its minimum value of \num{-0.14} between $\theta = \ang{150}$ and $\theta = \ang{180}$.
During the downwind half of the blade's rotation ($\ang{180} \leq \theta < \ang{360}$), there is a reversal between the suction and pressure sides of the blade.
A second counter-rotating leading-edge vortex now forms on the extrados of the blade's circular path ($\theta = \ang{240}$).
This vortex is less coherent and less strong than the one formed during the upwind.
It still yields a local maximum in the power coefficient around \num{0.1}.
The second leading edge vortex does not remain bound to the blade for long and before we reach $\theta = \ang{270}$, a layer of negative vorticity is entrained between the vortex and the blade surface.
The lift-off of the second leading edge vortex leads again to a drop in the power coefficient below \num{0}.
The power coefficient reaches a local minimum of \num{-0.07} at $\theta = \ang{329}$.
This phase position corresponds to the moment when the blade's angle of attack returns below its critical stall angle and flow reattachment is initiated.

Dynamic stall plays a central role in the temporal evolution of the power coefficient for vertical-axis wind turbines operating at low tip-speed ratios.
For all cases with $\lambda \leq 2.4$ in our configuration, we observe a succession of six characteristic stages, including attached flow,
shear layer growth,
upwind vortex formation,
upwind stall,
downwind stall,
and flow reattachment.
The wind turbine's performance varies significantly from one stage to the next.
Here, we focus on the characterisation of the temporal development of the flow structures during the full dynamic stall life cycle, their influence on the unsteady load response and turbine power production, and changes in the life cycle as a function of the tip-speed ratio.

\subsection{\label{subsec:PODres} Proper orthogonal decomposition and modal analysis}

\begin{figure}[tb]
\centering
\includegraphics{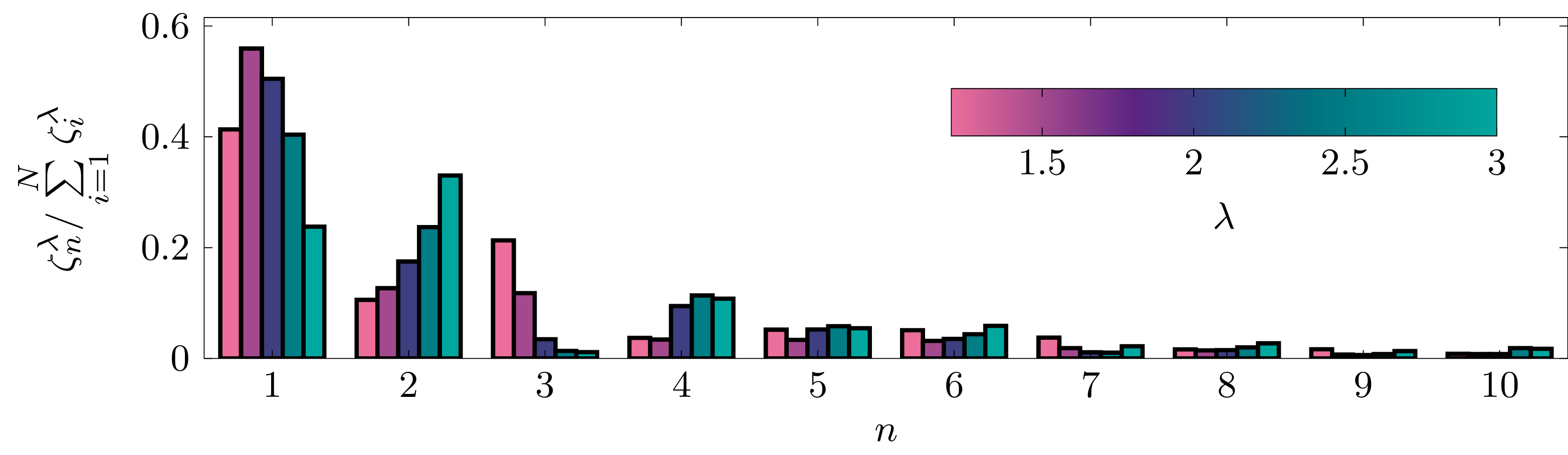}
\caption{Normalised eigenvalues associated with the first ten spatial POD modes for different tip-speed ratios calculated according to \cref{eq:zeta}.
}
\label{fig:res_PODlambda}
\end{figure}

\begin{figure}[tb]
\centering
\includegraphics{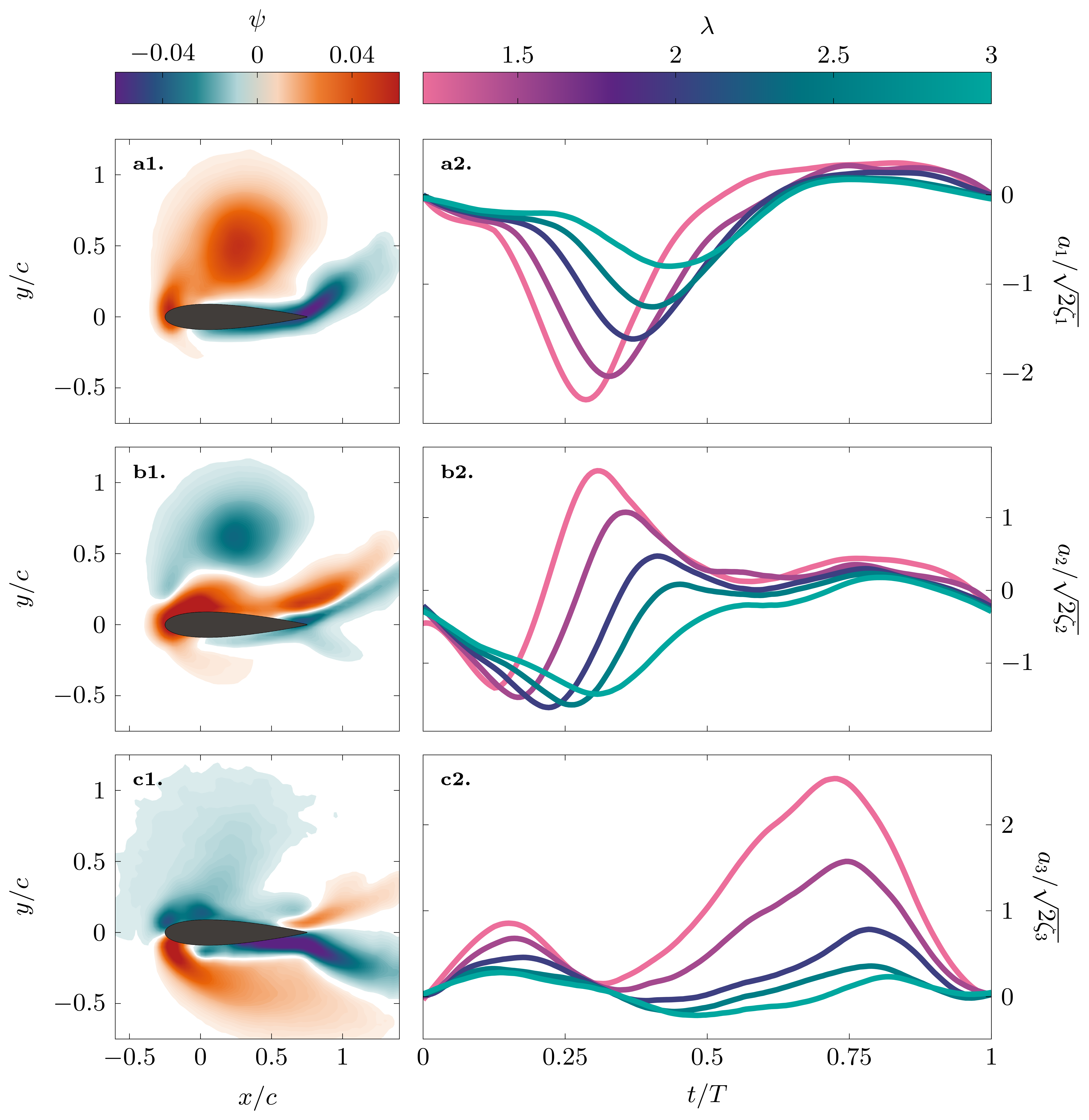}
\caption{First three spatial POD modes \subf{a1-c1} and the evolution of the corresponding time coefficients for tip-speed ratios $\lambda = 1.2, 1.5, 2.0, 2.5$ and $3.0$ \subf{a2-c2}.
}
\label{fig:res_PODpiv}
\end{figure}

We use the proper-orthogonal decomposition (POD) of the vorticity field to identify the energetically dominant spatial modes that characterise the flow around our vertical-axis wind turbine blade.
The POD time coefficients associated with the dominant spatial modes are examined to identify the characteristic dynamic stall timescales.
A single set of spatial eigenmodes is extracted from the ensemble of phase-averaged vorticity fields measured at different tip-speed ratios following the procedure described in \cref{subsec:POD}.

The eigenvalues $\kindex{\zeta}{n}^{\lambda}$ associated with the first ten spatial POD modes for different tip-speed ratios are presented in \cref{fig:res_PODlambda}.
The tip-speed ratio specific eigenvalues are calculated according to \cref{eq:zeta} and represent the energy contribution of the individual spatial modes in the vorticity field reconstruction for the specific tip-speed ratio.
The eigenvalues are normalised by the sum of the tip-speed-ratio-specific eigenvalues.
Overall, the energy contribution of the modes decreases rapidly with increasing mode number and the first three modes represent more than \SI{70}{\percent} of energy for all tip-speed ratios.
Analysing flow structures with POD requires striking a balance between conciseness and completeness.
Selecting too few modes means one might ignore a too large energetic portion of the flow, while selecting too many sacrifices the benefits of a low-order analysis.
Here, we focus on the first three spatial modes to analyse the temporal development of dominant flow features.

The first three spatial POD modes are presented in \cref{fig:res_PODpiv}.
The corresponding time coefficients are shown for tip-speed ratios $\lambda = 1.2, 1.5, 2.0, 2.5$ and $3.0$.
The time coefficients \kindex{a}{n} are normalised by the eigenvalues obtained from the stacked flow field POD (\cref{subsec:POD}).
The first spatial mode is characterised by the presence of a large dynamic stall vortex with a vortex centre located at mid-chord and half-chord length above the blade's surface.
Note that the modes are presented in the blade's frame of reference.
Everything that is above the blade corresponds to what happens on the intrados of the blade's rotational trajectory.
The temporal evolution of the corresponding time coefficient $\kindex{a}{1}$ shows a distinctive peak when the upwind dynamic stall vortex is prominent.
The amplitude of the peak increases with decreasing tip-speed ratio.
The increase in amplitude highlights a more dominant large-scale vortex for decreasing tip-speed ratio.
At lower tip-speed ratios, the maximum effective angle of attack increases (\cref{fig:vawtaero}) and the strength of the dynamic stall vortex increases \citep{LeFouest2022}.
The first POD mode and the corresponding time coefficient can be used to characterise the timescales related to the formation of the upwind dynamic stall vortex.

The interpretation of the second spatial POD mode is more intricate than the first one, as it seems to combine two consecutive events that occur in the flow.
The time coefficient $\kindex{a}{2}$ is negative at the start of the upwind and initially decreases for all cases ($0 \leq t/T < 0.25$).
During this time period, flow reversal is first observed on the suction side of the turbine blade or the intrados of the blade's rotational trajectory.
Once the blade exceeds its critical stall angle, the time coefficient $\kindex{a}{2}$ starts to increase, becomes positive, and peaks at the occurrence of vortex lift-off.
The increase in $\kindex{a}{2}$ can be interpreted as the transition from a strong accumulation of negative vorticity near the surface during flow reversal, to the growth of a large vortex that entrains positive vorticity near the surface that leads to separation of the main upwind stall vortex.
The positive peak of the second mode coefficient occurs at the same time as the negative peak of the first mode coefficient.
The first two modes do not form a classical mode pair, where energy is transferred from one to the other, they rather reinforce each other.
The second POD mode and the corresponding time coefficient can be used to characterise the transition from the accumulation of vorticity and growth of the shear layer to the roll-up of the shear layer into a large-scale upwind dynamic stall vortex.

The third spatial POD mode does not display the presence of a coherent flow structure but rather a state of massively separated flow.
The corresponding time coefficient $\kindex{a}{3}$ significantly increases immediately after the separation of the upwind stall vortex, which corresponds to the moment when the amplitude of the first time coefficient $\kindex{a}{1}$ decreases.
The third spatial mode represents the fully separated flow following vortex detachment.
The severity of the post-stall conditions increases with decreasing tip-speed ratio.
Low tip-speed ratio cases are characterised by the formation of a stronger dynamic stall vortex than intermediate and high tip-speed ratio cases and they experience more pronounced deep stall events.
The absence of a coherent flow structure in the third spatial mode explains the asymmetry in power production between the upwind and downwind halves of the rotation presented exemplarily in \cref{fig:res_DS}.
During upwind, a torque-producing coherent leading-edge vortex is formed.
During downwind, the flow is massively separated and the loads are characterised by drag excursions.

The time coefficients of the first three POD modes will be used next to identify the timing of the different dynamic stall stages.
The fourth POD mode, which plays a more prominent role than the third mode for higher tip-speed ratios where stall is less prominent, did not provide additional information for the identification of the stall stages.
It is included for reference in the supplementary material.

\subsection{\label{subsec:PODnet} Identification of dynamic stall stages using POD time coefficients}


\begin{figure}[tb]
\centering
\includegraphics{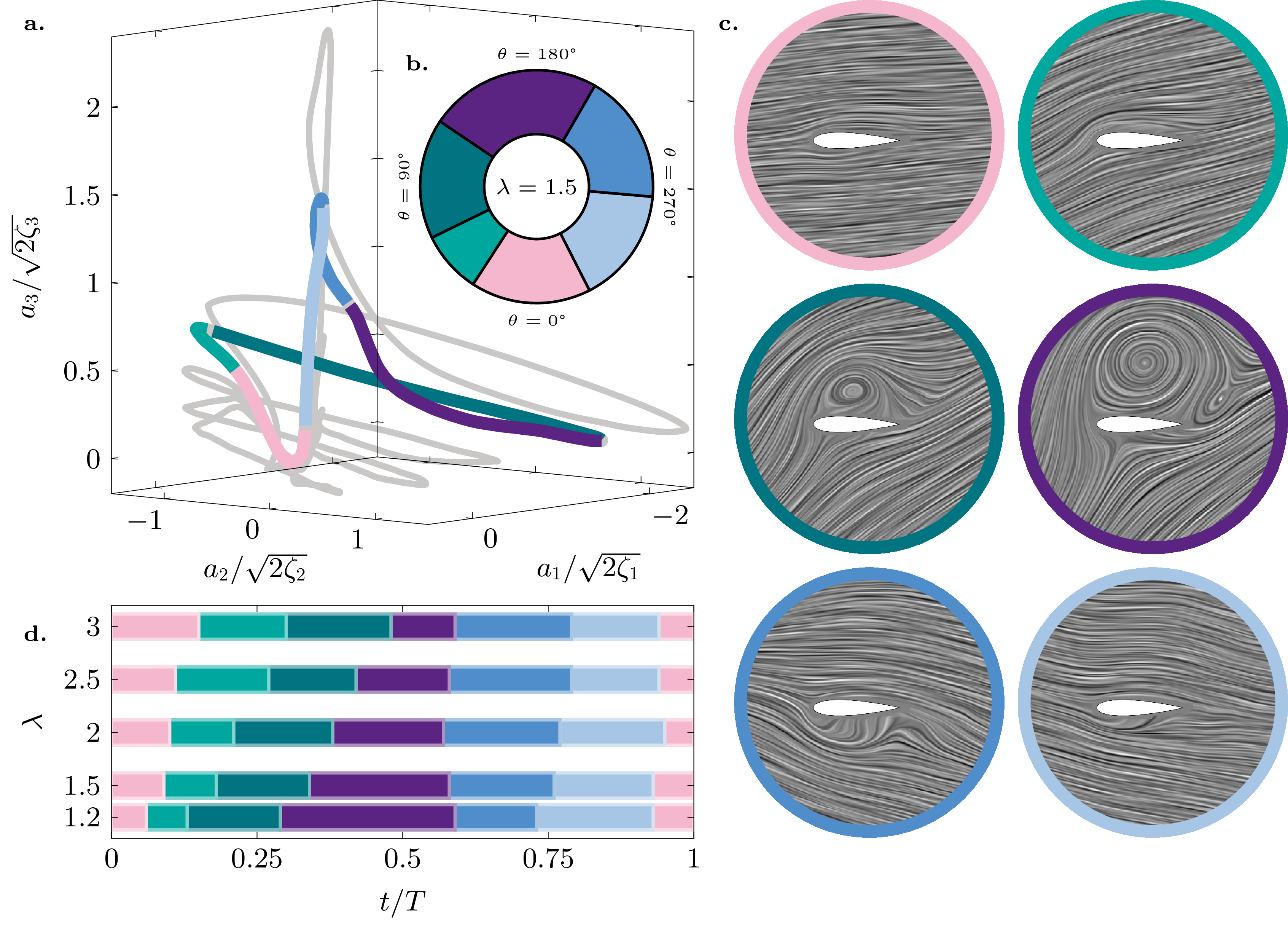}
\caption{\subf{a} Time coefficient parametric curve obtained from the stacked vorticity field POD (\cref{subsec:POD}).
The inset \subf{b} shows the phase map of the characteristic dynamic stall stages experienced by the wind turbine blade operating at tip-speed ratio $\lambda = 1.5$.
The stages are:
attached flow \protect\stage{stage1},
shear layer growth \protect\stage{stage2},
vortex formation \protect\stage{stage3},
upwind stall \protect\stage{stage4}, %
downwind stall \protect\stage{stage5},
and flow reattachment \protect\stage{stage6}.
\subf{c} Selected snapshots of the flow topology representing the characteristic stall stages obtained by the line integral convolution method.
\subf{d} Duration and timing of the dynamic stall stages for tip-speed ratio cases $\lambda = 1.2, 1.5, 2.0, 2.5$ and $3.0$.
}
\label{fig:resnet1}
\end{figure}

The trajectories in the feature space spanned by the POD time coefficients corresponding to the three dominant modes are shown in \cref{fig:resnet1}a for tip-speed ratio cases $\lambda = 1.2, 1.5, 2.0, 2.5$ and $3.0$.
The trajectory corresponding to the tip-speed ratio $\lambda = 1.5$ is colour-coded and used as an example to demonstrate our approach to identifying the timing of the various stages in the flow development.
The feature space representation highlights the interplay between the time coefficients.
We analyse the points of inflection of the trajectory.
A change in the relative importance of the time coefficient creates an inflection point on the trajectory in the feature space and relates to key events in the flow development.
The inflection points were systematically compared to the phase-averaged flow fields to determine their physical significance.
The results presented here are manually extracted and subject to the authors' discretion.
The inflection points were found by inspecting the three dimensional curves from various orientations.
Not all of them are visible in the two dimensional projection depicted in \cref{fig:resnet1}a.
Additional projections are included in the supplementary material.

We identify the timing and duration of six landmark stages:
attached flow (\stage{stage1}),
upwind shear layer growth stage (\stage{stage2}),
upwind vortex formation stage (\stage{stage3}),
upwind stalled stage (\stage{stage4}),
downwind stalled stage (\stage{stage5}),
and flow reattachment (\stage{stage6}).
A phase map for the six stages is shown for $\lambda = 1.5$ in \cref{fig:resnet1}b.
Visualisations of the flow topology characteristic of the individual stages are obtained by applying the line integral convolution method (LIC) to phase averaged flow field snapshots (\cref{fig:resnet1}c).
The timing and duration of the dynamic stall stages are summarised in \cref{fig:resnet1}d.

The same six stages occur for all tip-speed ratios, but the timing and duration of these stages vary.
The flow is attached when the blade enters the upwind part of the rotation for all cases.
First signs of stall development occur when the blade exceeds its critical stall angle, which happens later in the cycle for higher tip-speed ratios (\cref{fig:vawtaero}b).
Within less than a quarter of a rotation, a coherent upwind stall vortex has formed, which continues to grow in chord-wise and chord-normal direction until the vortex lifts off and opposite-signed vorticity is entrained between the vortex and the blade's surface.
The entrainment of secondary vorticity ultimately leads to the detachment of the stall vortex and the onset of upwind dynamic stall.
This process is similar to the vortex-induced separation observed on two-dimensional pitching and plunging airfoils \citep{Doligalski1994, Mulleners2012, Rival:2014bf}.

The upwind dynamic stall onset is identified in the POD feature space in \cref{fig:resnet1}a as the local maximum of $\kindex{a}{2}$ and the minimum of $\kindex{a}{1}$.
The time interval between the moment when the blade exceeds its critical stall angle and the onset of stall is defined as the upwind stall development.
We divide the upwind stall development into two parts in analogy with the flow development observed on pitching airfoils \citep{Mulleners2013, Deparday2019}.
The first part is characterised by the accumulation of bound vorticity and the thickening of the surface shear layer.
We call this stage the shear layer growth stage.
The shear layer growth stage (\stage{stage2}) takes up a larger portion of the cycle when the tip-speed ratio increases (\cref{fig:resnet1}d).
A kink in the feature space at the minimum of $\kindex{a}{2}$ and a local extremum of $\kindex{a}{1}$ marks the start of the roll-up of the shear layer into a coherent dynamic stall vortex.
This second part of the upwind stall development is characterised by the growth of the upwind dynamic stall vortex and is called the upwind vortex formation stage.
The vortex formation stage (\stage{stage3}) occupies approximately the same portion of the cycle for all tip-speed ratios but ends earlier in the cycle for lower tip-speed ratios (\cref{fig:resnet1}d).
Stall onset marks the end of the vortex formation stage and the start of the upwind stalled stage.

The next transition is marked by the switch of the leading edge stagnation point from the intrados of the blade path to the extrados, which corresponds respectively to the bottom and the top side of the blade in the snapshots presented in \cref{fig:vawtaero}c.
This transition is identified in the POD feature space as an inflection point in the gradient of $\kindex{a}{3}$ and occurs at approximately the same point in the cycle (\cref{fig:resnet1}d).
As the blade enters the downwind half of the rotation, the fully separated flow has no time to reattach to the blade even though the angle of attack is theoretically at \ang{0} when $\theta = \ang{180}$.
Furthermore, the blade's angle of attack varies more rapidly in the first part of the downwind than during the upwind.
There is insufficient time for a clear shear layer growth and vortex formation stage during the downwind, so we only consider an overall downwind stalled stage \citep{LeFouest2022}.
The end of the downwind stalled stage is marked by the maximum value of $\kindex{a}{3}$ and it is followed by the flow reattachment stage.
The flow reattaches to the blade for all cases in the last quarter of the turbine's rotation.

The characteristic upwind dynamic stall delay is the combined duration of the shear layer growth and the vortex formations stage.
Based on previous work, we expect that the duration of the stall delay expressed in convective time scales becomes independent of the kinematics for larger reduced frequencies \citep{LeFouest2021, Ayancik.2022}.
Generally, the non-dimensional convective time is obtained by dividing the physical time by $c/\U$, with $\U$ a constant characteristic velocity.
Here, the effective flow velocity experienced by the blade changes constantly and instead of using the blade velocity or the free-stream velocity to obtain a non-dimensional convective time, we opt here to take the constantly varying effective flow velocity into account and calculate the convective time perceived by the turbine blade using:
\begin{equation}
	\int_{0}^{t - \kindex{t}{ss}} \frac{\kindex{\U}{eff}(t)}{c} \mathrm{~d} t,
\end{equation}
where \kindex{U}{eff} is the effective flow velocity obtained from the vectorial combination of the blade velocity and the incoming flow velocity (\cref{fig:vawtaero}).

\begin{figure}[tb]
\centering
\includegraphics{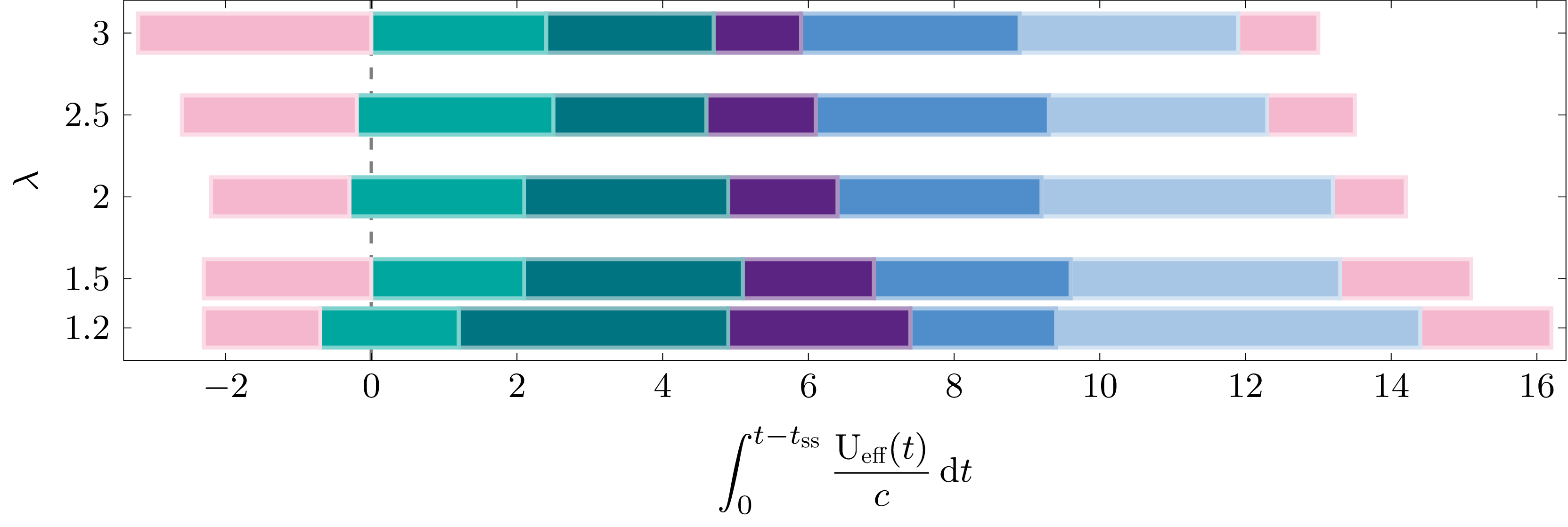}
\caption{Duration and timing of the dynamic stall stages for tip-speed ratio cases $\lambda = 1.2, 1.5, 2.0, 2.5$ and $3.0$ in terms of the convective time.
The stages are:
attached flow \protect\stage{stage1},
shear layer growth \protect\stage{stage2},
vortex formation \protect\stage{stage3},
upwind stall \protect\stage{stage4}, %
downwind stall \protect\stage{stage5},
and flow reattachment \protect\stage{stage6}.
}
\label{fig:res_scaled_stages}
\end{figure}

The temporal occurrence of dynamic stall stages scaled with the convective time is presented in \cref{fig:res_scaled_stages}.
Note that we subtract the time at which the static stall angle is exceeded \kindex{t}{ss} to realign subsequent stall events, in accordance with results presented in \cite{Buchner2018}.
When we express the upwind stall delay (\stage{stage2}+\stage{stage3}) in terms of convective times, the stall delay does not depend on the tip-speed ratio and converges to approximately \num{4.5} convective times, which matches standard vortex formation times found in literature \citep{Gharib1998,Dabiri2009,Dunne2016a,Kiefer.2022}.
The combination of the upwind and downwind stalled stages also reaches similar values for all tip-speed ratios, although the distribution between the two stages varies.
The flow reattachment timescale decreases slightly for increasing tip-speed ratios.
This trend may be due to the discrepancy between the effective flow velocity calculated using \cref{eq:Ueff} and the actual effective flow velocity acting on the turbine blade in the downwind half of the turbine rotation.
\Cref{eq:Ueff} neglects any influence of the induced velocity from shed vorticity, which is abundant in the downwind half of the turbine's rotation.
Overall, the convective time offers a promising scaling option for dynamic stall timescales on vertical axis turbines.
Further work can aim at modelling the influence of vorticity induced velocity to improve the scaling and implement universal timescales into dynamic stall modelling and control strategies for vertical-axis wind turbines.

\subsection{\label{subsec:loadsnet} Identification of dynamic stall stages using the aerodynamic loads}

\begin{figure}[tb]
\centering
\includegraphics{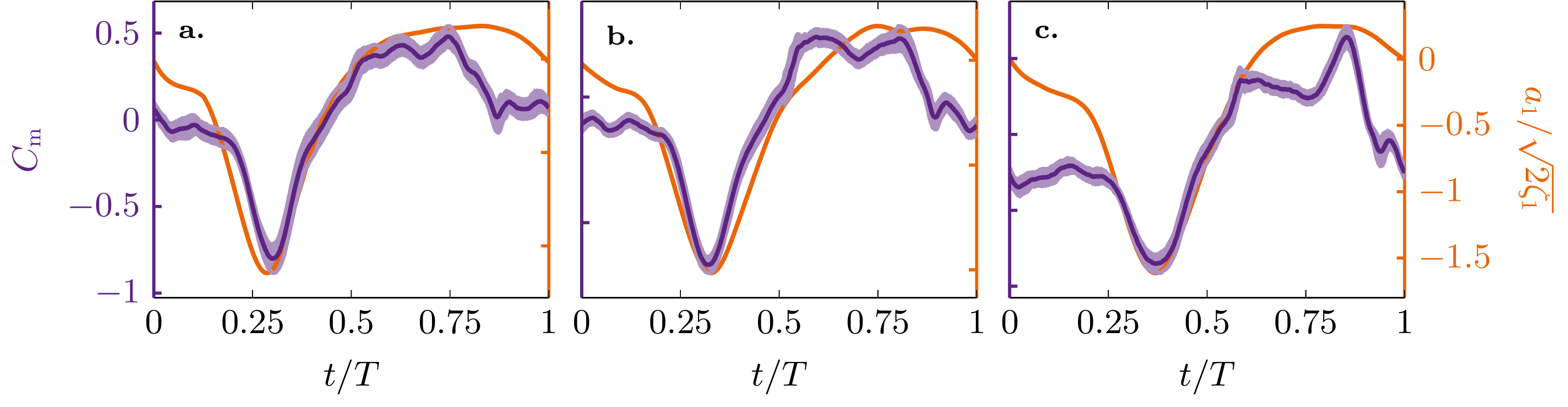}
\caption{Comparison of the temporal evolution of the first POD mode time coefficient $\kindex{a}{1}$ and the phase-averaged pitching moment coefficient for tip-speed ratio cases $\lambda = 1.2, 1.5, 2.0$.
}
\label{fig:res_a1Cmcomp}
\end{figure}

We cannot rely on flow measurements if we want to achieve real-time active flow control on a wind turbine blade.
The processing time of the flow measurement around the blade, capturing vortex formation or incoming gusts, for instance, is prohibitively high to actuate a response in time.
Finding patterns in the readily available aerodynamic loads that indicate the occurrence of stall stages is desirable.
This feasibility of replacing the time series of the POD coefficients with load measurement time series was inspired by the similarity between some features in the POD time coefficients and the temporal development of the aerodynamic loads.
In \cref{fig:res_a1Cmcomp}, we compare the temporal evolution of the time coefficient $\kindex{a}{1}$ corresponding to the first POD mode and the phase-averaged pitching moment coefficient for tip-speed ratios $\lambda = 1.2, 1.5, 2.0$.
The general shape of the two curves is in close agreement.
The timing of the distinctive peak displayed by the time coefficient $\kindex{a}{1}$ perfectly matches the timing of the negative pitching moment peak.
The pitching moment is highly sensitive to changes in the formation of the large-scale upwind dynamic stall vortex, which is characterised by the mode coefficient $\kindex{a}{1}$.
This further confirms our interpretation of the first POD mode and motivates us to extract the timing and duration of the dynamic stall stages based solely on the aerodynamic loads.

\begin{figure}[tb]
\centering
\includegraphics{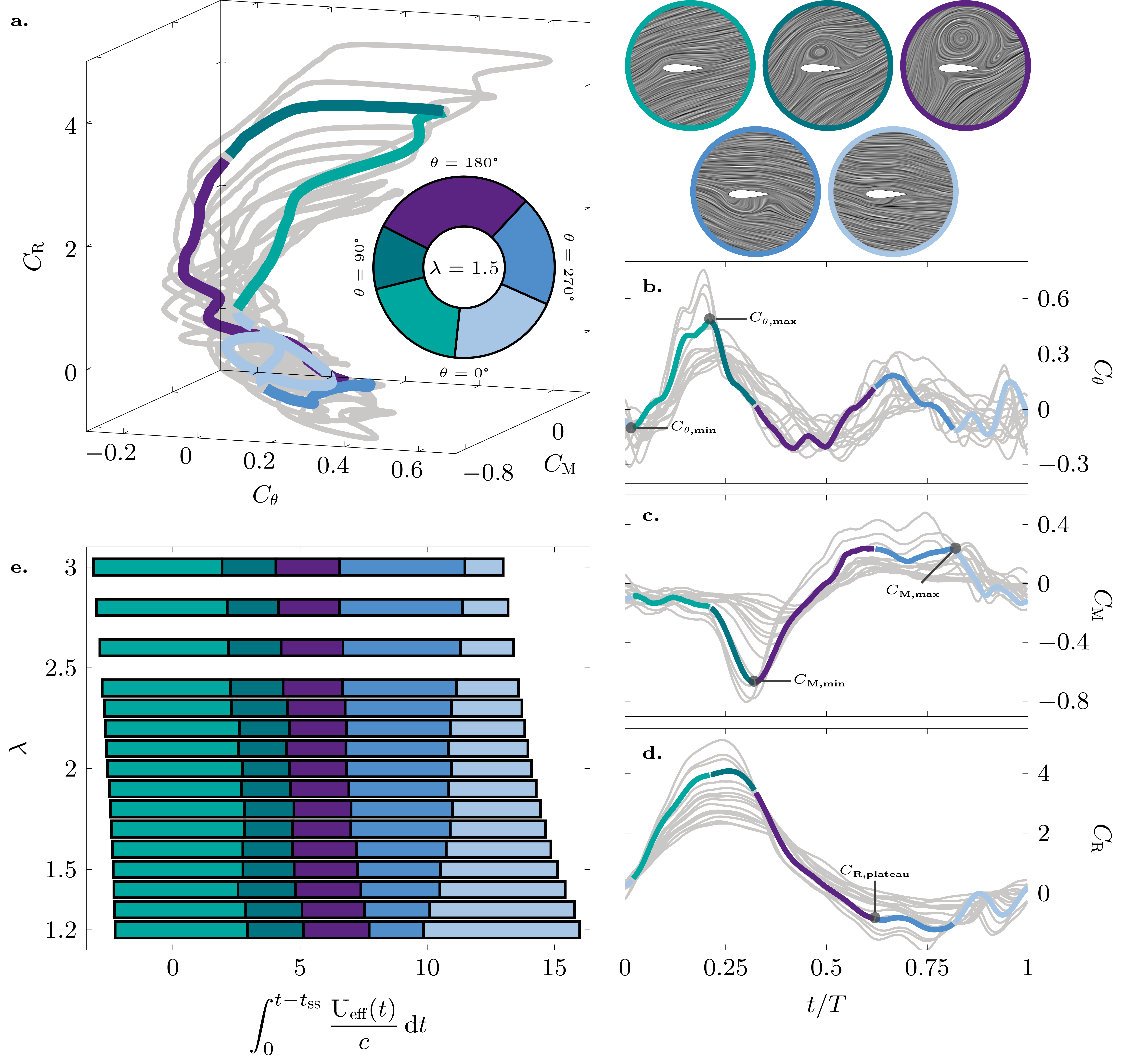}
\caption{\subf{a} Unsteady force parametric curve for tip-speed ratios $\lambda \in [1.2-3.0]$.
The inset shows the phase map of the characteristic dynamic stall stages experienced by the wind turbine blade operating at tip-speed ratio $\lambda = 1.5$.
The stages are:
attached flow and shear layer growth \protect\stage{stage2},
vortex formation \protect\stage{stage3},
upwind stall \protect\stage{stage4}, %
downwind stall \protect\stage{stage5},
and flow reattachment \protect\stage{stage6}.
\subf{b} Duration and timing of the dynamic stall stages retrieved from the unsteady loads for tip-speed ratio $\lambda \in [1.2-3.0]$.
Temporal evolution of phase-averaged unsteady \subf{c} tangential load coefficient \kindex{C}{$\theta$}, \subf{d} pitching moment coefficient \kindex{C}{M}, \subf{e} radiant force coefficient \kindex{C}{R} for tip-speed ratios $\lambda \in [1.2-3.0]$.
Selected snapshots of the flow topology representing the characteristic stall stages are repeated as reminders.
}
\label{fig:res_netloads}
\end{figure}

We analyse the dynamic stall timescales based on the development of the unsteady loads experienced by the wind turbine blade using a similar approach as for the POD time coefficients.
Our new feature space is built using the azimuthal force, radial force and pitching moment coefficients and analyse the measured trajectories in \cref{fig:res_netloads}a for 19 tip-speed ratio cases with $\lambda \in [1.2-3.0]$.
The trajectory corresponding to the tip-speed ratio $\lambda = 1.5$ is colour-coded and used as an example to demonstrate the automated identification of the dynamic stall stages from the unsteady loads.
Automated identification increases the robustness and the potential of using unsteady loads as the input for active flow control laws.
The points of inflection formed by the unsteady load trajectories are not as clearly defined as those in the POD time coefficient trajectory shown in \cref{fig:res_PODpiv}a, but we are able to identify five extrema that are systematically featured in the temporal development of the aerodynamic loads and that can be directly related to the changes in the dynamic stall life cycle.
These extrema were identified by closely analysing the interplay between flow structures developing around the turbine blade and abrupt changes in the load response.

The first extremum we identified is the tangential force minimum occurring shortly after the blade enters the upwind half of its rotation ($0 \leq t/T < 0.5$) (\cref{fig:res_netloads}b).
This time instant is followed by a strong increase in torque production, which corresponds to the shear-layer growth stage (\cref{fig:res_DS}, $\theta = \ang{45}$).
The end of the shear-layer growth stage coincides with the tangential force maximum occurring around $t/T = 0.25$.
The vortex formation stage is characterised by the convection of the upwind dynamic stall vortex from the leading edge towards the mid-chord position, causing a significant excursion of the pitching moment coefficient from just below \num{0} to its minimum value \kindex{C}{M,min} (\cref{fig:res_netloads}c).
The pitching moment minimum coincides with vortex separation and the beginning of the upwind stalled stage.
Vortex separation is followed by a significant loss in radial force that decreases until around $t/T = 0.6$, reaching a plateau that corresponds to the downwind stalled stage.
The transition from a sharp decrease to a plateau is identified using the function \textit{findchangepts} in MATLAB.
Lastly, the pitching moment coefficient reached a local maximum before returning to its initial value during the flow reattachment stage.
These extrema are 
used to identify the landmark dynamic stall stages for all tip-speed ratio cases $\lambda \in [1.2 - 3.0]$.

A phase map showing the temporal occurrence of dynamic stall stages identified using unsteady loads is shown as an inset in \cref{fig:res_netloads}a for $\lambda= 1.5$.
The stages we identify from loads are the same as those identified using the time coefficient of the vorticity field POD with one exception: the attached flow stage.
There is no clear indication of an attached stage in the unsteady load response.
The blade experiences an immediate load response to the increase of the effective angle of attack at the beginning of the blade's rotation, but the appearance of flow reversal and shear-layer growth is delayed to higher angles of attack.
The first stage (\stage{stage2}) shown in \cref{fig:res_netloads}e thus combines the attached flow stage and the shear layer growth stage.
The timing and duration of the following stages match those computed with the POD time coefficients.
These findings suggest that the aerodynamic loads contain sufficient information to identify whether a wind turbine blade is undergoing dynamic stall and which stage the flow development is currently in.

\subsection{\label{subsec:Comp} Stage-wise comparison of both identification methods}

\begin{figure}[tb]
\centering
\includegraphics{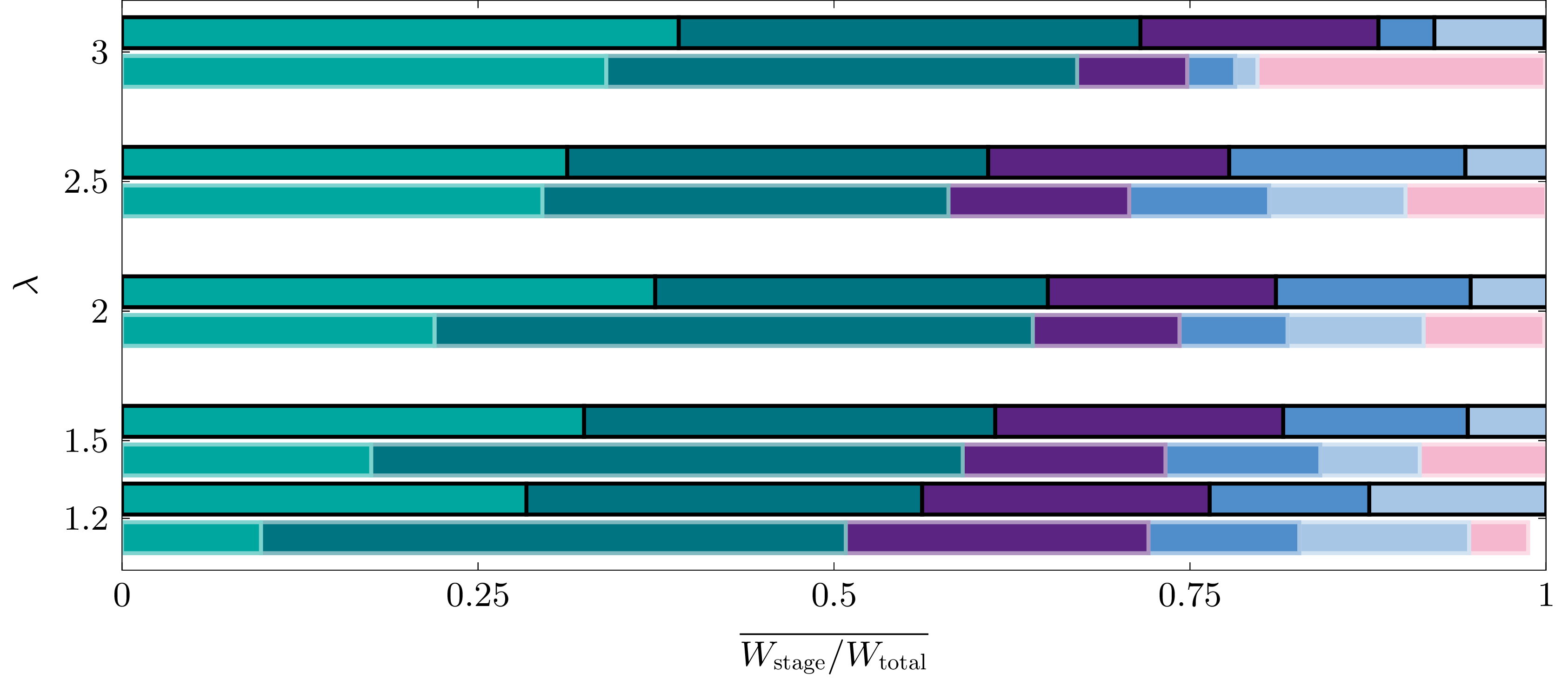}
\caption{Stage-wise contribution to the total work completed by the total aerodynamic force throughout the blade's rotation. The stage contributions are compared for stages identified either by using aerodynamic loads (dark contours) or by using POD time coefficients (light contours) for tip-speed ratios $\lambda = [1.2, 1.5, 2.0, 2.5\ \mathrm{and}\ 3.0]$.
The stages are:
attached flow \protect\stage{stage1},
shear layer growth \protect\stage{stage2},
vortex formation \protect\stage{stage3},
upwind stall \protect\stage{stage4}, %
downwind stall \protect\stage{stage5},
and flow reattachment \protect\stage{stage6}.
}
\label{fig:res_barComp}
\end{figure}

A direct comparison of both stage identification methods is desirable to assess the potential of unsteady loads to capture the timescales of flow development for analytical or control purposes.
To allow for a direct comparison between the stages computed from the aerodynamic loads and those obtained using the POD time coefficients, we calculated the work done by the total aerodynamic force in each stage for both methods.
The work done by the total aerodynamic force over a given stage is calculated with:
\begin{equation}
\kindex{W}{stage} = \overline{\kindex{F}{tot}} \kindex{\Theta}{stage} R
\end{equation}
where $\overline{\kindex{F}{tot}}$ is the mean total aerodynamic force experienced by the turbine blade over that stage and $\kindex{\Theta}{stage}$ is the angular distance covered by the blade throughout the stage.
This metric allows a pertinent comparison between the two stage identification methods because it accounts for both the duration of the stage and the magnitude of the loads experienced by the turbine blade during that stage.
A small error in the timescale of a given stage is enhanced if the stage is related to significant unsteady loads.
Alternatively, we could have chosen the blade's time-averaged power production over each stage as a metric, but we preferred the work done by the total force because it accounts for both the radial and tangential force components and has a high signal-to-noise ratio, even for high tip-speed ratio cases when the force magnitude is lower.

The stage-wise contributions to the work completed by the aerodynamic force are compared in \cref{fig:res_barComp} for stages identified using either the aerodynamic loads or the POD time coefficients for tip-speed ratios $\lambda = [1.2, 1.5, 2.0, 2.5\ \mathrm{and}\ 3.0]$.
To facilitate the comparison, the work completed in each stage \kindex{W}{stage} is normalised by the total work completed by the aerodynamic force throughout the blade's rotation $\kindex{W}{total} = \overline{\kindex{F}{tot}} 2 \pi R$.

Overall, most of the aerodynamic work is completed during the shear layer growth (\stage{stage2}) and the vortex formation stage (\stage{stage3}).
These stages are associated with the accumulation of vorticity close to the blade's surface, which creates significant suction over about one quarter of the turbine's rotation.
This trend is verified for both stage identification methods at all tip-speed ratios.
The combined contribution of the shear layer growth and the vortex formation stage is nearly independent of tip-speed ratio and reaches $0.6$ for all cases.
Both stage identification methods show excellent agreement.
The aerodynamic loads do not allow for a clear identification of the attached flow stage.
Instead, the attached flow stage is distributed across the flow reattachment and mainly the shear layer growth stages.
The work contribution of the attached flow stage is limited across tip-speed ratios.
The accurate capture of key timescales, such as the vortex formation time, is a satisfactory and promising result for the use of aerodynamic loads as a tool to identify flow development timescales and implement closed-loop control strategies.


\section{\label{sec:conclu}Conclusion}


We collected time-resolved velocity field and unsteady load measurements on a vertical-axis wind turbine operating at low tip-speed ratios $\lambda \in [1.2 - 3.0]$ to characterise the chain of events that leads to dynamic stall and to quantify the influence of the turbine operation conditions on the duration of the individual stages of the flow development.
The dominant flow features and their timescales were first analysed using a proper orthogonal decomposition of the stacked vorticity fields from different tip speed ratio conditions.
This procedure is known as a parametric modal decomposition and yields a single set of representative spatial modes and tip-seed ratio specific temporal coefficients and eigenvalues that can be directly compared to characterise the influence of the tip-speed ratio on the temporal development of the dominant spatial flow features.

Based on the POD time coefficients of the three most energetic modes, we were able to identify the timing and duration of six landmark dynamic stall stages: the attached flow, shear layer growth, vortex formation, upwind stall, downwind stall, and flow reattachment stage.
The duration of the attached flow increases with increasing tip-speed ratio as the blade exceeds its critical stall angle later for higher tip-speed ratios.
The combined duration of the shear layer growth and vortex formation stages represent the characteristic dynamic stall delay.
This dynamic stall delay is independent of tip-speed ratio for our experimental conditions and reaches a constant non-dimensional value of $4.5$ convective times, which corresponds to the typical dynamic stall delays found on non-rotating pitching and plunging airfoils at comparable reduced frequencies.
The upwind and downwind stall stages shorten with increasing tip-speed ratio as the blade spends less time at angles of attack beyond the critical stall angle.
Flow reattachment starts once the blade returns below its critical angle and full reattachment is reached within \numrange{3}{4} convective times for all cases studied.
The dynamic stall stages were also identified based solely on the measured aerodynamic loads.
The timing of the dynamic stall stages based on the loads agrees well with the timing based on the time coefficients of the first three modes of the POD of the vorticity field.

Our findings demonstrate that aerodynamic loads are suitable for analysing the timescales of flow development on a vertical axis wind turbine blade and can be used for flow control applications.
However, a control strategy should not necessarily aim at avoiding the onset of dynamic stall as the work completed by the aerodynamic forces during the upwind dynamic stall delay represents approximately \SI{60}{\percent} of the total work completed throughout the turbine's rotation.
Future work should focus on careful management of the dynamic stall vortex formation and shedding to exploit the work done by the vortex but avoid potential negative ramifications related to its shedding.

\begin{Backmatter}

\paragraph{Funding Statement}
Support from the Swiss national science foundation under grant number PYAPP2\_173652 is gratefully acknowledged.

\paragraph{Declaration of Interests}
The authors declare no conflict of interest.


\paragraph{Data Availability Statement}
The data used in this work will be made available upon request.



\bibliography{main.bib}

\end{Backmatter}

\newpage
\appendix
\section{Appendix}

\subsection{Vertical-axis wind turbine model}

A scaled-down model of a single-bladed H-type Darrieus wind turbine was mounted in the centre of the test section.
The turbine has variable diameter $D$ that was kept constant here at \SI{30}{\centi\meter}.
Up to three blades can be attached to the rotor arms through straight shafts that are held from the top.
Here, we used the single-blade configuration to focus on the flow development around the blade in the absence of interference from the wakes of other blades.
The turbine blade itself was 3D printed using a photosensitive polymer resin (Formlabs Form 2 stereolithography), sanded with very fine P180 grit paper and covered with black paint.
The blade has a NACA0018 profile with a span of $s=\SI{15}{\centi\meter}$ and a chord of $c=\SI{6}{\centi\meter}$, yielding a chord-to-diameter ratio of $c/D=\num{0.2}$.
The turbine blade is fully submersed and its top is located at $\kindex{h}{0} = \SI{12}{\centi\meter}$ below the water level.
This yields a Froude number $\textrm{Fr} = \frac{\Uinf}{\sqrt{g \kindex{h}{0}}} \in [0.26 - 0.65]$ for our tested range of tip-speed ratio $\lambda \in [1.2 - 3]$.

The turbine's compact geometry allowed for a blockage ratio of \SI{12.5}{\percent}, based on the ratio of the turbine's swept area, given by the blade's span times the rotor diameter, and the water channel's cross-section.
The blade is held by a cantilevered shaft such that there is no central strut interference with the flow.
At low tip-speed ratios, the effective blockage is closer to \SI{2.5}{\percent}, which is the blockage ratio calculated based on the ratio of the blade area, given by the blade's chord times its span, to the cross-sectional area.
Additionally, a \num{2.5} chord length distance to the water channel's side walls is also respected at all times.
Based on these observations, we consider the blockage and confinement effects small, and we did not apply any blockage correction to the force measurements \citep{Parker2017,Ross2020}.
This assumption is supported by a numerical simulation of a single-bladed wind turbine with a \SI{32}{\percent} blockage ratio conducted by \cite{SimaoFerreira2009}.

The turbine model is driven by a NEMA 34 stepper motor with a \ang{0.05} resolution for the angular position.
The rotational frequency was kept constant at \SI{0.89}{\hertz} to maintain a constant chord-based Reynolds number of $\kindex{\Rey}{c} = (\rho \omega R c)/\mu = \num{50000}$, where $\rho$ is the density and $\mu$ the dynamic viscosity of water.
To investigate the role of the tip-speed ratio in the occurrence of dynamic stall, we systematically vary the water channel's incoming flow velocity from \SIrange{0.14}{0.70}{\meter\per\second} to obtain tip-speed ratios ranging from \numrange{1.2}{6}.

\subsection{Load measurements}

The blade shaft was instrumentalised with twenty strain gauges forming five full Wheatstone bridge channels to record unsteady aerodynamic loads acting on the blade.
The strain gauges are powered and their output signal is amplified using an instrumentation amplifier with precision voltage reference placed on a printed circuit board that is mounted directly on the rotor arm.
The load cell was calibrated in situ using a fully orthogonal calibration rig and undergoing \num{1350} independent loading conditions.
A calibration matrix was obtained by performing a linear regression on the calibration measurements.
This matrix contained the \SI{95}{\percent} confidence interval of the coefficients allowing an estimation of the uncertainty related to numerous factors, including response linearity, hysteresis, repeatability, and measurement error.
The uncertainty was found to be below \SI{5}{\percent} for the shear force and pitching moment components.
A full description of the calibration procedure that includes the loading conditions, calibration matrix calculation, and error estimation can be found in the appendix of our previous work \citep{LeFouest2022}.

For each experiment, the wind turbine model starts at rest with the blade facing the incoming flow.
The turbine blade is then accelerated to its prescribed rotational speed.
After reaching the target rotational speed, we wait for five full turbine rotations before starting the load recordings.
Aerodynamic forces acting on the turbine blade are recorded at \SI{1000}{\hertz} for \num{100} full turbine rotations, then the blade is brought to rest.
The forces presented in this paper are the two shear forces applied at the blade's mid-span in the radial \kindex{F}{R} and azimuthal \kindex{F}{$\theta$} direction, and the pitching moment about the blade's quarter-chord \kindex{M}{1/4}.
The total force applied to the blade is computed by combining the two shear forces: $\kindex{F}{tot} = \sqrt{\kindex{F}{R}^2 + \kindex{F}{$\theta$}^2}$.
All force coefficients are non-dimensionalised by the blade chord $c$, the blade span $s$, and the blade velocity $\kindex{\U}{b} = \omega R$ such that:
\begin{equation*}
\kindex{C}{tot/R/$\theta$} = \frac{\kindex{F}{tot/R/$\theta$}}{0.5\rho \kindex{\U}{b}^2 sc}\quad.
\end{equation*}
The subscripts tot, R, or $\theta$ refer to the total force, the radial, or the tangential force component.
The centripetal force resulting from the turbine's rotation was experimentally measured by operating the wind turbine in air.
The added drag from the two splitter plates was measured and offset for all investigated tip-speed ratios by operating the wind turbine without the blade, where the two splitter plates were held by a small cylinder.
The influence of the centripetal force and the splitter plate are subtracted from the raw measurement data to isolate and compare the aerodynamic forces acting on the turbine blade.
A more detailed description of the measurement and modelling of the non-aerodynamic forces is included in the appendix of our previous work \citep{LeFouest2022}.
The presented force data was filtered using a second-order low-pass filter with the cut-off frequency at \SI{30}{\hertz}.
This frequency is multiple times larger than the pitching frequency and approximately \num{50} times larger than the expected post-stall vortex shedding frequency based on a chord-based Strouhal number of \num{0.2} \citep{Roshko1961}.

\subsection{Particle image velocimetry}

High-speed particle image velocimetry (PIV) was used to measure the flow field around the wind turbine blade.
A dual oscillator diode-pumped ND:YLF laser ($\lambda = \SI{527}{\nano\meter}$) with a maximum pulse energy of \SI{30}{\milli\joule} and a beam splitter were used to create two laser sheets from opposite sides of the channel.
The light sheets were oriented horizontally at mid-span of the turbine blade.
A high-speed camera with a sensor size of \SI{1024x1024}{\pixel} (Photron Fastcam SA-X2) and a spinning mirror apparatus were installed below the channel to capture the flow around the blade.
The spinning mirror apparatus comprises two rotating and one stationary mirror, all oriented with a \ang{45} angle with respect to the horizontal plane.
The two moving mirrors rotate about the same axis of rotation and at the same frequency as the wind turbine.
One of the mirrors is placed at the same radius as the model blade, such that it keeps the blade in the field of view of the camera at all times.
The spinning mirror apparatus allows us to measure the velocity field around the blade with a higher spatial resolution and without scarifying the temporal resolution.
The field of view is \SI{2.5x2.5}{c} centred around the blade.
The acquisition frequency is \SI{1000}{\hertz}.
The images were processed following standard procedures using a multi-grid algorithm \citep{Raffel2007}.
The final window size was \SI{48x48}{\pixel} with an overlap of \SI{75}{\percent}.
This yields a grid spacing or physical resolution of $\SI{1.7}{\milli\meter} = 0.029c$.
A window overlap above \SI{50}{\percent} was selected to minimise spatial averaging of the velocity gradients by the interrogation window following \cite{Richard.2006, kindlerAperiodicityFieldFullscale2011}.
The out-of-plane vorticity component was calculated from the in-plane velocity components using a central difference scheme. 
Outliers were identified and removed based on a vorticity threshold and median filter. 
A missing-vector rate of \SI{2}{\percent} was estimated accross the different measurements series. 

Due to the unsteady effective flow condition experienced by the blade throughout each rotation, the particle displacement between consecutive camera images varies greatly.
To have sufficient particles displacement between correlated images and similar relative displacements errors at all azimuthal rotor positions, we have used a variable image skip.
We have skipped a different number of particle images depending on the expected flow velocities at a given location and experimental conditions. 
The number of images skipped between the correlated images was determined based on the expected particle displacement assuming that the flow velocity away from the blade matches the effective flow velocity.
When the blade is at an azimuthal location where the expected effective velocity is low, more images are skipped and the time delay between the correlated images is increased.  
When the blade is at an azimuthal location where the expected effective velocity is high, no or fewer images are skipped and the time delay between the correlated images is decreased. 
At all positions, we aimed for a minimal partical displacement of \SI{12}{px} when the flow moves with the nominal effective velocity. 

\subsection{Phase-averaging}

Load measurements were obtained and phase-averaged over \num{100} wind turbine revolutions for all tip-speed ratio cases.
Phase-averaging involves splitting the phase space into \num{540} bins that cover \ang{0.67} without overlap.
For each bin, we calculate the mean performance value and its standard deviation.
This method allows us to visualise the mean performance of the turbine blade at \num{540} phase positions and the corresponding cycle-to-cycle variations of the performance.
The number of bins was selected to be large enough such that sufficient data points lie in each bin and small enough to reduce the smoothening of the data.
The acquired data and bin size yielded converged average and first-order statistical metrics of the unsteady aerodynamic loads experienced by the blade.

In addition to the load measurements, we conducted time-resolved velcoity field measurements using PIV for the tip-speed ratios $\lambda \in \{1.2, 1.5, 2, 2.5, 3\}$.
For every tip-speed-ratio, we collected a total of \num{21838} instantaneous particle images at an acquisition rate of \SI{1000}{\hertz}.
The images have been pairwisely correlated using variable image gaps to obtain instantaneous velocity fields. 
These instantaneous velocity fields were then divided into \num{200} bins that each cover \ang{1.8} without overlap.
The same azimuthal resolution is used for all tip-speed-ratios. 
The azimuthal resolution for the velocity field phase-averaging is smaller than for the load phase-averaging.
The velocity fields were recorded for \num{21} turbine rotations.
The load data was recorded for \num{100} rotations. 
The rotational velocity of the turbine was \SI{0.89}{\hertz} for all tip-speed-ratios as the tip-speed-ratio was altered by changing the incoming flow velocity.

Proper orthogonal decoposition is applied on the resulting phase-averaged flow fields.
This operation drastically decreases the computational cost, which scales with the number of snapshots squared.

\begin{figure}[tb!]
	\centering
	\includegraphics{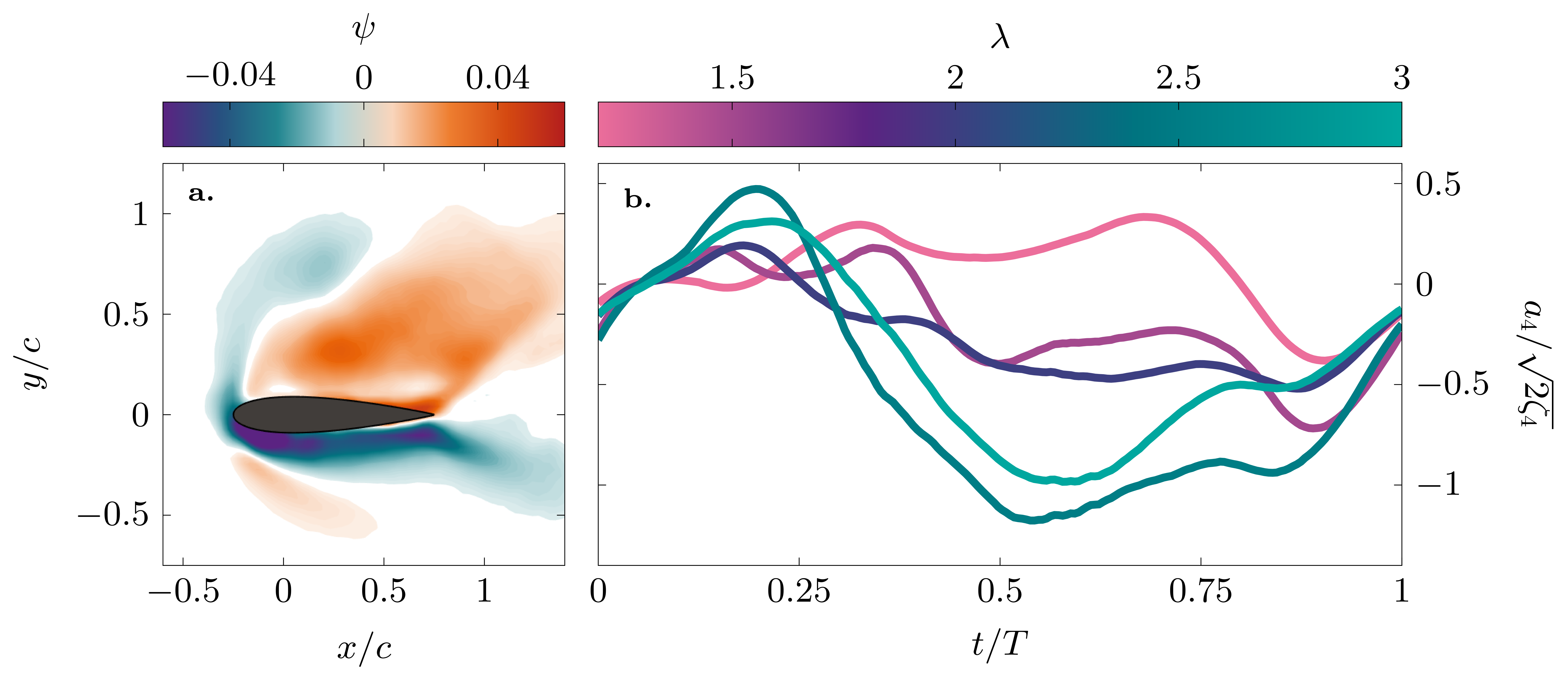}
	\caption{\subf{a} Fourth spatial POD mode and \subf{b} the evolution of the corresponding time coefficients for tip-speed ratios $\lambda = 1.2, 1.5, 2.0, 2.5$ and $3.0$.}
	\label{fig:mode4}
	\end{figure}

\subsection{Fourth POD mode}

The first three spatial POD modes and the corresponding time coefficients for tip-speed ratios $\lambda = 1.2, 1.5, 2.0, 2.5$ and $3.0$ are included in the main manuscript.
They have been used to identify the dynamic stall development stages.

The fourth POD mode, which plays a more prominent role than the third mode for higher tip-speed ratios where stall is less prominent, did not provide additional information for the identification of the stall stages.
It is shown in \cref{fig:mode4} for reference.

The fourth spatial mode represents the second leading-edge vortex that forms in the downwind half.
The time coefficients indicating the contribution of the fourth POD mode reach higher values for the higher tip-speed ratios ($\lambda = 2$, $2.5$, and $3$) than for the low tip-speed ratio ($\lambda = 1.2$ and $1.5$).
For low tip-speed ratios, the downwind half is mostly dominated by deep post-stall conditions and fully separated flow, whereas the higher tip-speed ratio cases are generally able to recover and form a second coherent leading-edge vortex, yielding an improved efficiency.

\subsection{Alternative viewing angles of the time coefficient parametric curves}

\begin{figure}[tb!]
\centering
\includegraphics{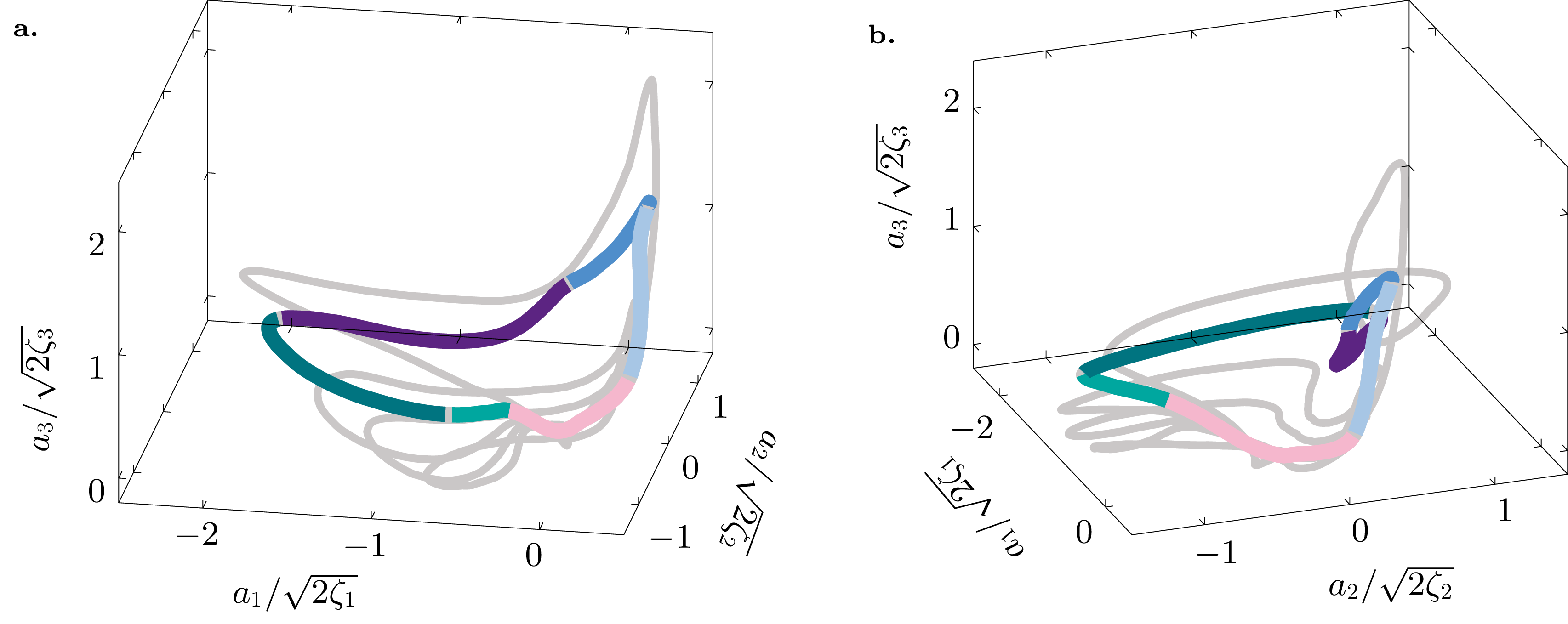}
\caption{Time coefficient parametric curve obtained from the stacked
vorticity field POD, similar to figure 6a in the manuscript, but
under two different orientations.
The curve corresponding to tip-speed ratio $\lambda=1.5$ is highlighted.
}
\label{fig:resnet_view}
\end{figure}

The stage detection is based on inflection points in the three-dimensional parametric curves.
Depending on the orientation of the plot, not all inflection points are clearly
visible.
Additional views are presented in \cref{fig:resnet_view}.

\end{document}